\documentclass[sigconf]{acmart}

\makeatletter
\@ifundefined{Bbbk}{}{
  
}
\makeatother

\PassOptionsToPackage{table}{xcolor}  
\usepackage{rotating}
\usepackage{subfigure}
\usepackage{tikz}
\usepackage{tikz-qtree}
\usepackage{pgfplots}
\usepackage{pgfplotstable}
\usepackage[hypcap=true]{subcaption}
\usetikzlibrary{patterns}
\usepackage{tabularx}
\usepackage{makecell}
\usepackage{booktabs}
\usepackage{bm}
\usepackage{enumerate, enumitem}
\usepackage{amsmath}
\usepackage{amssymb}
\usepackage{amsthm}
\usepackage{multirow}
\usepackage[ruled,linesnumbered,vlined,noend]{algorithm2e}
\usepackage{cleveref}
\usepackage{graphicx}
\usepackage{xcolor} 
\usepackage{colortbl} 
\usepackage{pifont}
\usepackage{setspace}
\usepackage{float}
\definecolor{tabcolor}{RGB}{217, 235, 246}
\definecolor{stdcolor}{gray}{0.3}  

\pgfplotsset{compat=1.18}

\newtheorem{problem}{Problem}
\AtBeginDocument{%
  }

\copyrightyear{2025}
\acmYear{2025}
\setcopyright{acmlicensed}\acmConference[SIGIR '25]{Proceedings of the 48th International ACM SIGIR Conference on Research and Development in Information Retrieval}{July 13--18, 2025}{Padua, Italy}
\acmBooktitle{Proceedings of the 48th International ACM SIGIR Conference on Research and Development in Information Retrieval (SIGIR '25), July 13--18, 2025, Padua, Italy}
\acmDOI{10.1145/3726302.3730263}
\acmISBN{979-8-4007-1592-1/2025/07}
\author{Yue Zhang}
\orcid{0009-0009-5199-7799}
\affiliation{%
  \institution{The Chinese University of Hong Kong, Shenzhen}
  \city{Shenzhen}
  \country{China}}
\email{yuezhang5@link.cuhk.edu.cn}

\author{Yankai Chen}
\orcid{0000-0001-5741-2047}
\affiliation{%
  \institution{University of Illinois Chicago}
  \city{Chicago}
  \country{United States}
}
\email{yankaichen@acm.org}

\author{Yingli Zhou}
\orcid{0009-0008-5630-6822}
\affiliation{%
  \institution{The Chinese University of Hong Kong, Shenzhen}
  \city{Shenzhen}
  \country{China}
}
\email{yinglizhou@link.cuhk.edu.cn}

\author{Yucan Guo}
\orcid{0009-0007-0125-4490}
\affiliation{%
  \institution{Institute of Computing Technology, Chinese Academy of Sciences}
  \city{Beijing}
  \country{China}
}
\email{guoyucan23z@ict.ac.cn}

\author{Xiaolin Han}
\orcid{0000-0002-4347-1692}
\affiliation{%
  \institution{Northwestern Polytechnical University}
  \city{Xi'an}
  \country{China}
}
\email{xiaolinh@nwpu.edu.cn}

\author{Chenhao Ma}
\orcid{0000-0002-3243-8512}
\affiliation{%
  \institution{The Chinese University of Hong Kong, Shenzhen}
  \city{Shenzhen}
  \country{China}
}
\email{machenhao@cuhk.edu.cn}
\authornote{Corresponding author}

\begin{document}
\begin{spacing}{0.965}

\title{UTCS: Effective Unsupervised Temporal Community Search with \\  Pre-training of Temporal Dynamics and Subgraph Knowledge}


\newcommand\model{UTCS}

\newcommand{\guo}[1]{{\textcolor{blue}{\textsf{#1}}}}
\renewcommand{\shortauthors}{Yue Zhang et al.}
\newcommand{\zhou}[1]{\textcolor{magenta}{[zhou: #1]}}
\begin{abstract}
  In many real-world applications, the evolving relationships between entities can be modeled as temporal graphs, where each edge has a timestamp representing the interaction time. 
  As a fundamental problem in graph analysis, {\it community search (CS)} in temporal graphs has received growing attention but exhibits two major limitations: (1) Traditional methods typically require predefined subgraph structures, which are not always known in advance. (2) Learning-based methods struggle to capture temporal interaction information. 
To fill this research gap, in this paper, we propose an effective \textbf{U}nsupervised \textbf{T}emporal \textbf{C}ommunity \textbf{S}earch with
pre-training of temporal dynamics and subgraph knowledge model (\textbf{\model}).
\model~contains two key stages: offline pre-training and online search.
In the first stage, we introduce multiple learning objectives to facilitate the pre-training process in the unsupervised learning setting.
In the second stage, we identify a candidate subgraph and compute community scores using the pre-trained node representations and a novel scoring mechanism to determine the final community members.
Experiments on five real-world datasets demonstrate the effectiveness. 

\end{abstract}

\ccsdesc[500]{Computing methodologies~Neural networks}
\ccsdesc[300]{Mathematics of computing~Graph algorithms}

\keywords{Community Search, Temporal Graph Mining, Hawkes Process}
\maketitle

\begin{figure*}[t]
    \centering
    \setlength{\abovecaptionskip}{-0.01cm}
    \setlength{\belowcaptionskip}{-0.01cm}
    \includegraphics[width=0.97\linewidth]{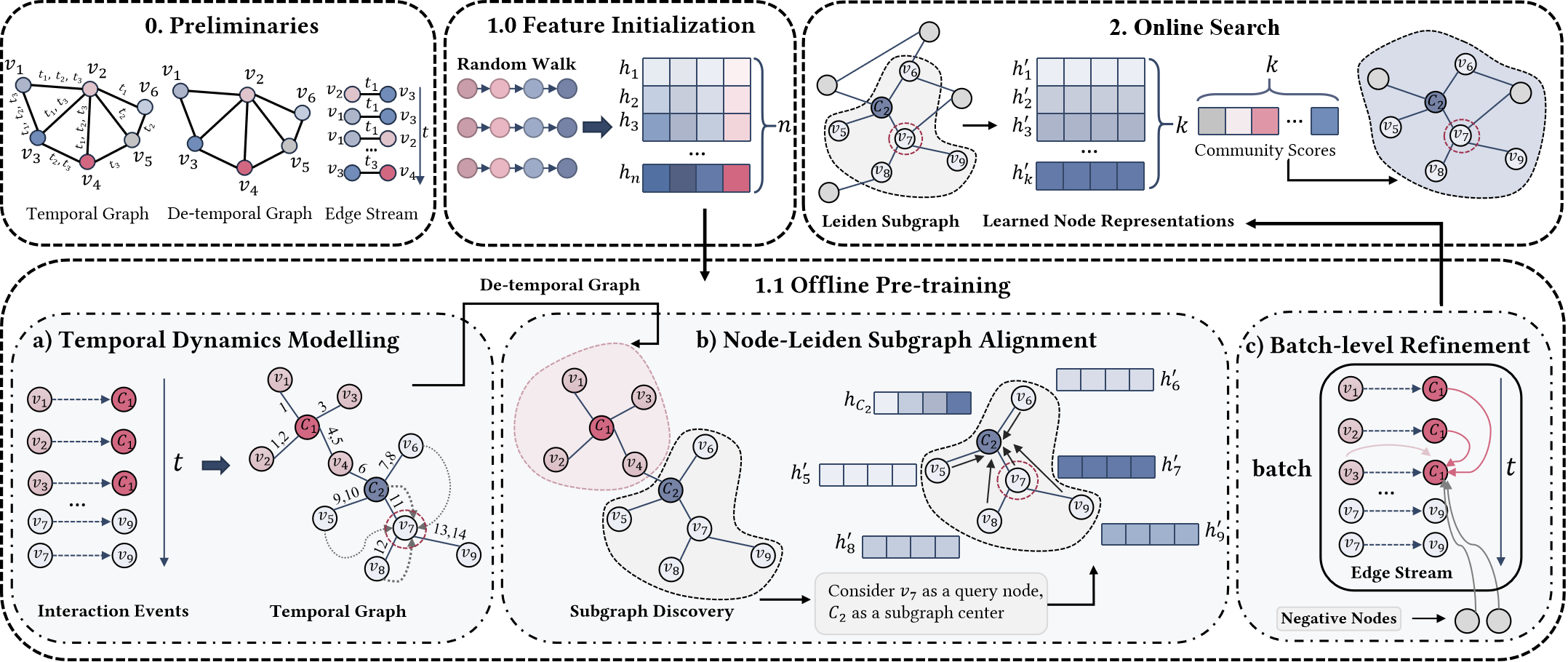}
    \caption{The architecture of our model.}
    \label{fig:model}
\end{figure*}
\section{Introduction}
\label{intro}
Graphs are widely utilized in various real-life fields \cite{zeng2024distributed, zeng2024efficient, luo2023efficient, luo2024efficient}, such as social networks, biological graphs, and financial graphs. Community Search (CS), as an important graph analytical problem, has attracted significant attention recently \cite{10.1007/s00778-019-00556-x}. 
Given a graph and a set of query nodes, the community search problem aims to find a cohesive and dense subgraph containing the query nodes~\cite{sozio2010community}. 
CS is widely used in applications like social recommendation, research community detection, and fraud group identification, inspiring numerous specialized approaches.
Existing methods can be broadly classified into two categories: traditional CS algorithms~\cite{akbas2017truss,yuan2017index, fang2017effective,chen2018exploring, zhou2023influential,zhou2024efficient} and learning-based CS models~\cite{gaoICSGNNLightweightInteractive2021, jiangQueryDrivenGraphNeural2022, hashemi2023cs, wangEfficientUnsupervisedCommunity2024, wangNeuralAttributedCommunity2024}.

Traditional CS algorithms rely on predefined subgraph patterns, such as $k$-core~\cite{cui2014local, sozio2010community,chen2021efficient}, $k$-truss~\cite{akbas2017truss, huang2015approximate}, and $k$-clique~\cite{cui2013online,yuan2017index,zhou2024counting}. These models impose strict topological constraints on the community, which may not always align with the structures of real-world communities. 
For instance, the $k$-core-based models assume that each node within the community has a degree of at least $k$, which may not always hold in reality. 
In contrast, learning-based CS models like QD-GNN~\cite{jiangQueryDrivenGraphNeural2022} and TransZero~\cite{wangEfficientUnsupervisedCommunity2024} adapt flexibly to graph structures without predefined patterns: QD-GNN separately models queries and communities via GNNs, while TransZero uses Graph Transformers to capture node-community proximity.

Despite the success of learning-based CS studies, most are designed for static graphs and overlook crucial \textbf{temporal interaction information}. In real-world temporal graphs~\cite{chu2019online}, such as business collaborations, two parties may establish relationships during specific periods. Identifying communities in such graphs is crucial for many real-world applications and has garnered growing research attention.
While temporal graphs can be treated as static for input into static models, this often leads to sub-optimal results.
For example, a recent learning-based work~\cite{hashemi2023cs} considers the graph within each time window as static, failing to capture temporal interactions within the time window. 
Moreover, most learning-based methods heavily rely on costly labeled data. This highlights the urgent need for an \textbf{unsupervised} approach that captures temporal information for effective community search.
To date, there is no existing learning-based approach for temporal community search in \textit{the unsupervised setting}.

{\bf Our solution.} To address the aforementioned limitations, we have developed an effective model, namely \textbf{U}nsupervised \textbf{T}emporal \textbf{C}ommunity \textbf{S}earch with
pre-training of temporal dynamics and subgraph knowledge (\textbf{\model}). 
\model~incorporates temporal and structural information at both local and global scales. As a learning-based method, it operates without requiring a predefined subgraph structure. 
Our model has two phases: offline pre-training and online search. During pre-training, we introduce the multiple learning objectives through three components:
a temporal dynamics learner based on the Hawkes process to model interaction patterns, a subgraph alignment mechanism leveraging Student's t-distribution and KL divergence to capture structural dependencies, and a batch-wise embedding refinement to reconstruct adjacency relations.
This phase enables unsupervised and label-free training. 
In the online search phase, we first locate candidate subgraphs, then leverage learned node representations to compute community scores and employ a dedicated matching mechanism to predict the final community structure.
Our main contributions are as follows.
\begin{itemize}[leftmargin=*]
    \item We propose a learning-based model for the temporal CS problem and, to the best of our knowledge, this is the first unsupervised approach to tackle it.
    \item Our model follows a two-step framework with modules designed to capture temporal, local, and global structural information. Additionally, it employs a local search approach to significantly reduce time and space complexity.
    \item Extensive experiments on five real-world datasets demonstrate the effectiveness of the model, achieving an average improvement of 60.44\% in F1-score compared to the latest competitors.
\end{itemize}
\section{Problem Formulation}
\label{sec:definitions}
In this paper, we consider an undirected temporal graph $\mathcal{G(V, E)}$, where $\mathcal{V}$ and $\mathcal{E}$ denote the sets of nodes and edges, respectively. Let $|\mathcal{V}|=n$ and $|\mathcal{E}|=m$. Each edge $e\in\mathcal{E}$ is a triplet $(u,v,t)$ with $u, v \in \mathcal{V}$ and $t\in\mathbb{N}$ representing the interaction timestamp between $u$ and $v$.
Note that $(u,v,t_1)$ and $(u,v,t_2)$ are considered as two edges when $t_1\neq t_2$. We introduce the concept of temporal neighbors $N_{(u, t)}$, where $N_{(u, t)} = \{w | (u,w,t_w)\in \mathcal{E} \text{ and } t_w<t\}$. 
Figure \ref{fig:model} preliminaries illustrate a sample
temporal graph $\mathcal{G}$ with 6 vertices and 17 temporal edges.
In addition, we define the concept of the de-temporal graph, a version of the temporal graph $\mathcal{G}$ with timestamps ignored as $\mathcal{G'(V', E')}$, where $\mathcal{E}' = \{(u,v)|\exists (u,v,t)\in \mathcal{E}\}$ and $|\mathcal{E}'| = \bar{m}$. 
Figure \ref{fig:model} preliminaries show a de-temporal graph $\mathcal{G'}$.
%
\begin{problem}[Temporal Community Search (TCS) \cite{hashemi2023cs}]
Given a temporal graph $\mathcal{G}$ and a set of query nodes $\mathcal{Q} \subseteq \mathcal{V}$, the TCS problem aims to return a subgraph (i.e., community) $\mathcal{T}_{\mathcal{Q}}$ containing $\mathcal{Q}$.
\end{problem}
In this paper, we focus on the learning-based approach to solve the TCS problem.
For a set of query nodes $\mathcal{Q} \subseteq \mathcal{V}$, and its corresponding community $\mathcal{T}_{\mathcal{Q}}$, the nodes in $\mathcal{V}$ can be classified into two categories based on the indicator function $f$ below, i.e., $\forall v \in \mathcal{V}$: $f(v) = 1$ if $v \in \mathcal{T}_{\mathcal{Q}}$; otherwise, $f(v) = 0$.
Specifically, our objective is to train a neural network $\mathcal{M}$ to identify communities in a temporal graph.

\section{Our Methodology}
\subsection{Overview}
In this section, we explain the details of \model. 
%
%
As depicted in Figure~\ref{fig:model}, 
\model~consists of two phases: offline pre-training and online search. 

%
\subsection{Offline Pre-training}
To better align with real-world communities, we address the TCS problem by training a neural network to model both temporal interaction information and structural relationships.
\subsubsection{\textbf{Temporal Dynamics Modeling}}
To establish a good starting point, we initialize node representations using node2vec~\cite{grover2016node2vec}. 
To incorporate temporal information, we model the temporal graph as a chronological order of edge streams, as shown in Figure~\ref{fig:model} preliminaries. 
Then, we resort to the Hawkes process~\cite{hawkes1971spectra,zuoEmbeddingTemporalNetwork2018}, where past interactions temporarily boost the likelihood of future events with exponentially decaying influence.
%
Inspired by~\cite{zuoEmbeddingTemporalNetwork2018, liuDEEPTEMPORALGRAPH2024}, we construct a temporal loss function based on the Hawkes process. This approach allows us to capture the influence of past events on current interactions, providing a comprehensive representation of temporal information. The temporal loss term can be formulated as:
\begin{equation}\label{4}
    L_{tmp} = -\log \sigma(\lambda_{(u,v,t)}) - \sum_{i=1}^l\mathbb{E}_{\bar{h}_i\sim  P_n(v)}\log \sigma(-\lambda_{(u,\bar{h}_i,t)}),
\end{equation}
where $\sigma$ is the sigmoid activation function and $\lambda_{(u,v,t)}$ is the conditional intensity function of $v$ joining source node $u$'s neighborhood. We sample negative nodes $\bar{h}$ with probability proportional to their degree $P_n(v) \propto d_v^{3/4}$~\cite{mikolov2013distributed}, using a fixed sample size of $l$.

\subsubsection{\textbf{Node-Leiden Subgraph Alignment}}
To capture the relationship between nodes and communities, we apply the Leiden algorithm~\cite{traag2019louvain} to obtain a set of Leiden subgraphs $\{\mathcal{L}_{i}\}$, and refine the node representations by incorporating the community structure captured by these subgraphs. The probability $q_{(v, i,t)}$ of node $v$ belonging to a Leiden subgraph $\mathcal{L}_i$ at time $t$ is computed using Student’s t-distribution~\cite{van2008visualizing} based on embedding proximity. We derive the target distribution $p_{(v, i,t)}$ by squaring and normalizing the assignment probabilities to reinforce high-confidence predictions. Finally, we introduce the KL divergence~\cite{kullback1951information} loss term to align the real-time assignment distribution with the target distribution:
\begin{equation}
    L_{node} = \sum_{\mathcal{L}_i}p_{(v,i,t)}\log p_{(v,i,t)}/q'_{(v,i,t)}.
\end{equation}
Probability $q'_{(v, i,t)}$ is adjusted by updated node embeddings. This loss ensures proper node alignment with their respective subgraphs.

\subsubsection{\textbf{Batch-Level Embedding Refinement}}
Existing temporal graph methods often process edge streams in batches, overlooking structural connectivity. Inspired by~\cite{simclr,zhang2024geometric}, we propose a batch-level refinement that reconstructs adjacency relationships via node embeddings.
Specifically, similar to the computation of temporal conditional intensity in Eq.~(\ref{4}), we retrieve the historical neighbors $N_{(u,t)}$ as positive nodes and sample some non-neighbors $\overline{N}_{(u,t)}\sim P_n{(u)}$ as negative nodes.
By maximizing the cosine similarity between positive pairs and minimizing it for negative pairs, we capture local structural information. The loss term is thus defined as follows:
\begin{equation}
    \begin{gathered}
    L_{u} = -\log \frac{\exp \left( \cos(\boldsymbol{z}_u^t, \boldsymbol{z}_h^t)/T \right)}{\exp \left( \cos(\boldsymbol{z}_u^t, \boldsymbol{z}_h^t)/T \right) + \sum_{\bar{h} \in \overline{N}_{(u,t)}} \exp \left( \cos(\boldsymbol{z}_u^t, \boldsymbol{z}_{\bar{h}}^t)/T \right)} \\
    L_{batch} = \frac{1}{|N_{(u,t)}|+1} \sum_{h \in N_{(u,t)} \cup \{v\}}L_u ,
\end{gathered}
\end{equation}
where $\boldsymbol{z}^t_u$ and $\boldsymbol{z}^t_v$ are the temporal embeddings of source node $u$ and target node $v$ at current time $t$, respectively. $\cos(\boldsymbol{z}_u^t, \boldsymbol{z}_h^t)$ measures the cosine similarity between the anchor $\boldsymbol{z}^t_u$ and a positive sample $\boldsymbol{z}^t_h$, with temperature $T=0.5$ controlling the distribution sharpness.

By integrating the three modules, the complete community search loss function is formulated as:
$L = L_{tmp} + L_{node} + L_{batch}$.
\subsection{Online Search Phase}
\label{online search}
After offline pre-training, we move to an online search phase where nodes are embedded into a shared space, and proximity indicates similarity.  
A candidate $v$'s community score $s_v$ reflects its likelihood of belonging to the target community based on its distance to the query nodes $\mathcal{Q}$. The Expected Community Score Gain (ECSG) measures the gap between the scores of community nodes and the expected scores of random nodes~\cite{wangEfficientUnsupervisedCommunity2024}. 
To avoid costly global scoring, we adopt a local search strategy that reduces computation while maintaining comparable performance.
\begin{algorithm}
\footnotesize
\caption{Online Search}
\label{alg:local_search}
\SetAlgoLined
\KwIn{The query nodes $\mathcal{Q}$, de-temporal graph $\mathcal{G}'$ and offline pre-training network $f(\cdot)$.}
\KwOut{The predicted community $\mathcal{T}_{\mathcal{Q}}$.}
Initialize $\mathcal{D} \gets \mathcal{Q}, \{\mathcal{C}_i\} \gets \tt Leiden(\mathcal{G}')$\;
\ForEach{$\mathcal{C}_i$ with $\mathcal{C}_{i} \cap \mathcal{Q} \neq \emptyset$}{
        $\{\mathcal{C}_{i,j}| 1 \leq j \leq k \} \gets \tt TopKSimC(\mathcal{C}_i)$\;
        $\mathcal{D} = \mathcal{D} \cup \mathcal{C}_i \cup \mathcal{C}_{i,1} \cup \cdots \cup  \mathcal{C}_{i,k}$}
$z_{\mathcal{Q}} \gets \frac{\sum_{q \in \mathcal{Q}}f(q)}{|\mathcal{Q}|}$\;
\lFor{$v \in \mathcal{D}$}{
    $s_v \gets \frac{z_{\mathcal{Q}}\cdot f(v)}{\|z_{\mathcal{Q}}\|\cdot\|f(v)\|}$
}
Initialize $\mathcal{T}_{\mathcal{Q}} \gets \mathcal{Q},  max\_score \gets -\infty$\;
$\mathcal{S}=\{s_v|v \in \mathcal{D}\}$\;
\While{$\mathcal{|T_{Q}| < |D|}$}{
$u \gets \arg\max_{v \in \mathcal{D}\setminus \mathcal{T_{Q}}}{s_v}$\;
\uIf{${\tt ECSG}(\mathcal{S, T_{Q}}\cup \{u\}, \mathcal{G'}) > max\_score$}{
$ max\_score \gets {\tt ECSG}(\mathcal{S, T_{Q}}\cup \{u\}, \mathcal{G'})$\;
$\mathcal{T}_{\mathcal{Q}} = \mathcal{T}_{\mathcal{Q}} \cup \{u\}$
}
\lElse{
Break
}
}
\Return{$\mathcal{T_{Q}}$}\;
\end{algorithm}

The details are outlined in Algorithm \ref{alg:local_search}. Given query nodes $\mathcal{Q}$, the de-temporal graph $\mathcal{G}'$, and the pre-trained neural network, it outputs the predicted community $\mathcal{T}_{\mathcal{Q}}$ containing $\mathcal{Q}$.
We first partition $\mathcal{G}'$ via the Leiden algorithm to identify subgraphs ${\mathcal{C}_i}$ containing $\mathcal{Q}$ (lines 1-2), retrieve $k$ similar subgraphs with \texttt{TopKSimC}, and merge them into the search space $\mathcal{D}$ (lines 3-4).
Next, the community scores are computed for nodes in $\mathcal{D}$ based on cosine similarity with $\boldsymbol{z}_{\mathcal{Q}}$ (lines 5-6).
 Starting with $\mathcal{T}_{\mathcal{Q}}$ as $\mathcal{Q}$ and $max\_score$ as negative infinity, we iteratively add the node with the highest score in $\mathcal{D}\setminus \mathcal{T_{Q}}$ that increases the ECSG (lines 7-13).
 The search concludes by returning the final community $\mathcal{T}_{\mathcal{Q}}$. 


\section{Experiments}
\label{sec:ex}
\definecolor{c1}{RGB}{205,68,50} 
\definecolor{c2}{RGB}{62,134,181} 
\definecolor{c3}{RGB}{255,181,73} %
\definecolor{c4}{RGB}{183,131,175} 
\definecolor{c5}{RGB}{115,107,157} 

\definecolor{c6}{RGB}{205,68,50} 
\definecolor{c7}{RGB}{62,134,181} 
\definecolor{c8}{RGB}{239,145,99} 
\definecolor{c9}{RGB}{33,26,62} 
\definecolor{c10}{RGB}{208,108,157} 

\definecolor{c11}{RGB}{33,26,62} 
\definecolor{c12}{RGB}{208,108,157} 
\definecolor{c13}{RGB}{209,103,129}
\definecolor{c14}{RGB}{113,140,184}

\definecolor{c15}{RGB}{217, 235, 246}
\definecolor{c16}{RGB}{193, 207, 204}
\definecolor{c17}{RGB}{202, 206, 227}
\definecolor{c18}{RGB}{234, 170, 194}
\definecolor{c19}{RGB}{182, 226, 234}
\definecolor{c20}{RGB}{247, 244, 225}

\pgfplotstableread[row sep=\\,col sep=&]{
datasets & QDGNN & TransZero & \model \\
2 & 644 & 462  & 396 \\ 
4 & 10884 & 734 & 402  \\ 
6 & 15136 & 732 & 448  \\ 
8 & 100 & 832 & 1300  \\ 
10 & 100 & 832 & 1386 \\ 
}\memorysize

\pgfplotstableread[row sep=\\,col sep=&]{
datasets  & QD-GNN & COCLEP & TransZero & CS-TGN & \model \\
2  & 644 & 458& 462 & 602&396 \\ 
4  & 10884 & 7780 & 734 & 3480&402 \\ 
6  & 15136 & 1408 & 732 & 1666&448 \\ 
8  & 100 & 1140 & 832 & 100&1300 \\ 
10  & 100 & 1530 & 832 & 100&1386  \\ 
}\lmemorysize

\pgfplotstableread[row sep=\\,col sep=&]{
datasets & QTCS& QD-GNN & COCLEP & TransZero & CS-TGN & \model \\
2 & 115.8237 & 0.0086 & 0.0836& 0.0027 & 0.0287&0.0014 \\ 
4 & 26.9671 & 0.1254 & 3.6812 & 0.0277 & 0.0280&0.0701 \\ 
6 & 0.0837 & 0.5440 & 0.4564 & 0.0882 & 0.0106&0.0566 \\ 
8 & 0.7799 & 0 & 4.0976 & 0.2176 & 0&0.0736 \\ 
10 & 2.9276& 0 & 2.1804 & 0.3759 & 0&0.1475  \\ 
}\queryt

\pgfplotstableread[row sep=\\,col sep=&]{
data & QTCS & \model \\ 
2 & 112.2491 & 0.022  \\ 
4 & 27.6361 & 1.5224 \\ 
6 & 0.0362 & 0.1065  \\ 
8 & 2.2893 & 1.4536  \\ 
10 & 3.422 & 2.787  \\ 
}\grainedt



\pgfplotstableread[row sep=\\,col sep=&]{
data & QTCS & \model  \\ 
2 & 115.8237 & 0.0014  \\ 
4 & 26.9671 & 0.0701  \\ 
6 & 0.0837 & 0.0566  \\ 
8 & 0.7799 & 0.0736 \\ 
10 & 2.9276 & 0.1475  \\ 
}\initialt
\subsection{Setup}
\textbf{Datasets.}
\begin{table}[ht]
    \caption{Datasets statistics.}
    \setlength{\abovecaptionskip}{-2cm}
    \setlength{\belowcaptionskip}{-2cm}
    \footnotesize
    \begin{tabular}{c|c|c|c|c|c}
        \toprule
        Datasets&$|\mathcal{V}|$&$|\mathcal{E}|$&$|\mathcal{E'}|$&$t_{max}$&\#$\mathcal{C}$\\
        \midrule
        School~\cite{mastrandrea2015contact} & 327 & 188,508 & 5,802 & 7,375&9\\
        Brain~\cite{preti2017dynamic} & 5,000 & 1,955,488 & 1,751,910 & 12&10\\
        Patent~\cite{hall2001nber} &  12,214 & 41,916& 41,915 & 891&6\\
        arXivAI~\cite{wang2020microsoft} & 69,854 & 699,206 & 699,198 & 27&5\\
        arXivCS~\cite{wang2020microsoft} & 169,343 & 1,166,243 & 1,166,237 & 29&40\\
        \bottomrule
    \end{tabular}
    \label{tab:datasets}
\end{table}
We use five real-world temporal graphs across various domains~\cite{liuDEEPTEMPORALGRAPH2024}, each with ground-truth communities, e.g., academic collaboration networks and biological networks. 
\begin{table*}[t]
    \caption{Temporal Community Search. (The best and second best results are marked in bold and underlined respectively; ``OOM'' denotes the cases out-of-memory; ``-'' denotes the result is not available; ``< 0.01'' denotes the value is less than 0.01).}
    \label{tab:weak-connectivity}  
    \resizebox{\textwidth}{!}{
    \begin{tabular}{c|ccc|ccc|ccc|ccc|ccc|c}
        \toprule
        \multirow{2}{*}{Algorithms} & \multicolumn{3}{c|}{School} & \multicolumn{3}{c|}{Brain}& \multicolumn{3}{c|}{Patent} & \multicolumn{3}{c|}{arXivAI} & \multicolumn{3}{c|}{arXivCS} & \multirow{2}{*}{Rank} \\
        \cmidrule(lr){2-4} \cmidrule(lr){5-7} \cmidrule(lr){8-10} \cmidrule(lr){11-13}\cmidrule(lr){14-16}
        & Jaccard & F1-Score & NMI & Jaccard & F1-Score & NMI & Jaccard & F1-Score & NMI & Jaccard & F1-Score & NMI& Jaccard & F1-Score & NMI& \\
        \midrule
    {\tt MPC} &  0.0676 & 0.1267 & 0.0602
    & - & - & -
    & - & - & -
    & - & - & -
    & - & - & -& 7/7/7\\
    {\tt QTCS} &  0.7118& 0.8316 & 0.6198
     & 0.1590 & 0.2743 & 0.0529
     & 0.0205& 0.0402 & 0.0108
     & $<0.01$& $<0.01$ & $<0.01$
     & $<0.01$& $<0.01$ & $<0.01$ &5/5/4 \\
     \midrule
{\tt QD-GNN} 
&  0.75 ±  \textcolor{stdcolor}{0.31}
&  0.80 ± \small \textcolor{stdcolor}{0.29} 
&  0.68 ± \small \textcolor{stdcolor}{0.34}
 &  0.20 ± \small \textcolor{stdcolor}{0.07} 
 &  0.33 ± \small \textcolor{stdcolor}{0.08} 
 &  0.06 ± \small \textcolor{stdcolor}{0.05}
 & \textbf{ 0.24} ± \small \textcolor{stdcolor}{0.01} 
 & \textbf{ 0.38} ± \small \textcolor{stdcolor}{0.02} 
 &  0.05 ± \small \textcolor{stdcolor}{0.01}
 & OOM & OOM & OOM
 & OOM & OOM & OOM& 3/3/3\\
 {\tt  COCLEP} 
 & \underline{0.90} ± \small \textcolor{stdcolor}{0.02} 
 &  \underline{0.94} ± \small \textcolor{stdcolor}{0.01} 
 &  \textbf{0.87} ± \small \textcolor{stdcolor}{0.02}
 &  0.14 ± \small \textcolor{stdcolor}{0.04} 
 &  0.24 ± \small \textcolor{stdcolor}{0.05} 
 &  0.05 ± \small \textcolor{stdcolor}{0.04}
 &  \underline{0.19} ± \small \textcolor{stdcolor}{0.00} 
 &  0.30 ± \small \textcolor{stdcolor}{0.00} 
 &  \textbf{0.07} ± \small \textcolor{stdcolor}{0.00}
 &  \underline{0.12} ± \small \textcolor{stdcolor}{0.00} 
 &  \underline{0.19} ± \small \textcolor{stdcolor}{0.00} 
 &  \underline{0.04} ± \small \textcolor{stdcolor}{0.00}
 &  \underline{0.06} ± \small \textcolor{stdcolor}{0.00} 
 &  \underline{0.10} ± \small \textcolor{stdcolor}{0.00} 
 &  \underline{0.04} ± \small \textcolor{stdcolor}{0.00}
& 2/2/2\\
{\tt  TransZero} 
&  0.14 ± \small \textcolor{stdcolor}{0.00}
&  0.25 ± \small \textcolor{stdcolor}{0.00}
&  0.01 ± \small  \textcolor{stdcolor}{0.00}
&  \underline{0.24} ± \small \textcolor{stdcolor}{0.00}
&  \underline{0.38} ± \small \textcolor{stdcolor}{0.00}
&  \underline{0.11} ± \small \textcolor{stdcolor}{0.00}
 & 0.14 ± \small \textcolor{stdcolor}{0.00}
 & 0.24 ± \small \textcolor{stdcolor}{0.00}& $<0.01$ 
 & 0.09 ± \small \textcolor{stdcolor}{0.00}& 0.17 ± \small \textcolor{stdcolor}{0.00}& $<0.01$
 & 0.02 ± \small \textcolor{stdcolor}{0.00} & 0.03 ± \small \textcolor{stdcolor}{0.00} & $<0.01$ & 4/4/5
 \\
 {\tt  CS-TGN} &  0.69 ± \small \textcolor{stdcolor}{0.01} & 0.78 ± \small \textcolor{stdcolor}{0.01} &  0.63 ± \small \textcolor{stdcolor}{0.01}& 0.12 ± \small \textcolor{stdcolor}{0.01} 
& 0.21 ± \small \textcolor{stdcolor}{0.01} 
& 0.04 ± \small \textcolor{stdcolor}{0.01}
& 0.08 ± \small \textcolor{stdcolor}{0.02} 
& 0.14 ± \small \textcolor{stdcolor}{0.04} 
& 0.04 ± \small \textcolor{stdcolor}{0.01}
& OOM & OOM & OOM
 & OOM & OOM & OOM
& 6/6/6\\
     \midrule
     \rowcolor{tabcolor} 
        {\tt  \model} &  \textbf{0.93} ± \small \textcolor{stdcolor}{0.01} 
        &  \textbf{0.95} ± \small \textcolor{stdcolor}{0.01}
        &  \underline{0.83} ± \small \textcolor{stdcolor}{0.00}
         &  \textbf{0.32} ± \small \textcolor{stdcolor}{0.00}
         &  \textbf{0.49} ± \small \textcolor{stdcolor}{0.00} 
         &  \textbf{0.18} ± \small \textcolor{stdcolor}{0.00}
         &  0.18 ± \small \textcolor{stdcolor}{0.01} 
         &  \underline{0.31} ± \small \textcolor{stdcolor}{0.00}
         &  \underline{0.06} ± \small \textcolor{stdcolor}{0.00}
         &  \textbf{0.20} ± \small \textcolor{stdcolor}{0.00}
         &  \textbf{0.33} ± \small \textcolor{stdcolor}{0.00} 
         &  \textbf{0.10} ± \small \textcolor{stdcolor}{0.00}
        &  \textbf{0.12} ± \small \textcolor{stdcolor}{0.00} 
        &  \textbf{0.22} ± \small \textcolor{stdcolor}{0.00}
        & \textbf{0.07} ± \small \textcolor{stdcolor}{0.00}
        & 1/1/1
     \\
            \bottomrule
    \end{tabular}%
    }
\vspace{-0.2in}
\end{table*}
The datasets statistics are summarized in Table \ref{tab:datasets}. $|\mathcal{V}|$ is the number of nodes, and $|\mathcal{E}|$ denotes the number of temporal edges, $|\mathcal{E}'|$ represents the number of edges in the de-temporal graph by ignoring the temporal information, and $t_{max}$ denotes the number of different timestamps. \#$\mathcal{C}$ denotes the number of communities for each graph. Our code is available at \url{https://github.com/zyxxxx2/UTCS}.

\textbf{Baselines.}
Existing CS methods are categorized into traditional and learning-based.
We compare two traditional methods, {\tt MPC}~\cite{qin2019mining} and {\tt QTCS}~\cite{lin2024qtcs}. For the learning-based approaches, we compared three representative static CS methods and one temporal CS method, i.e., {\tt TransZero}~\cite{wangEfficientUnsupervisedCommunity2024}, {\tt COCLEP}~\cite{liCOCLEPContrastiveLearningbased2023}, {\tt QD-GNN}~\cite{jiangQueryDrivenGraphNeural2022} and {\tt CS-TGN}~\cite{hashemi2023cs}.

\textbf{Effectiveness Metrics and Implementation Details.}
In this paper, we mainly focus on the F1-score, Normalized Mutual Information (NMI)~\cite{danon2005comparing}, and Jaccard similarity (JAC)~\cite{zhang2020seal}, with higher values indicating better results. Models are trained for 200 epochs with embedding dimension 128, $T=0.5$, historical neighbors $h=3$, negative samples $\bar{h}=3$, learning rate 0.01, $k=2$, and batch size 1024. 
The final results are averaged over five runs. 
The baseline hyperparameters are set according to their respective original papers.
\subsection{Detailed Analysis of UTCS}
\label{ablation}
$\bullet$ \textbf{Effectiveness Evaluation.}
We compare {\tt \model} with baselines across five datasets by generating 100 random queries per dataset, averaging F1, JAC, and NMI over five runs.
As shown in Table~\ref{tab:weak-connectivity}:
(1) Traditional CS methods struggle with temporal communities. Specifically, {\tt MPC} often returns empty results, and {\tt QTCS} performs poorly on large datasets like arXivAI and arXivCS;
(2) Learning-based methods generally outperform traditional ones, suggesting real-world communities differ from predefined structures;
(3) Among learning-based models, {\tt \model} consistently achieves the best results, improving F1-score by 60.44\% over the second-best model, {\tt COCLEP}. 

$\bullet$ \textbf{Online Query Time.}
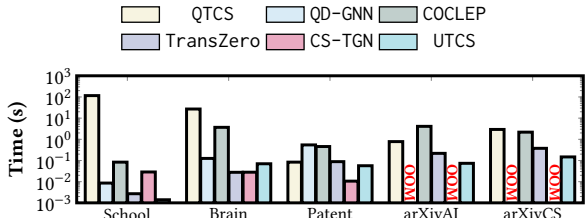
\begin{figure}[h]
    \centering
    \setlength{\abovecaptionskip}{-0.02cm}
    \setlength{\belowcaptionskip}{-0.1cm}

    \quad \ref{eff_role}\\
    \begin{tikzpicture}[scale=0.45]
        \begin{axis}[
            ybar=0.5pt,
            bar width=0.4cm,
            width=0.93\textwidth,
            height=0.3\textwidth,
            xtick=data,	
            xticklabels={\huge School,\huge Brain,\huge Patent,\huge arXivAI, \huge arXivCS},
            legend style={at={(0.5,0.98)},
            anchor=north,legend columns=3,
            draw=none},
            legend image code/.code={
                \draw [#1, line width=0.5pt] (0cm,-0.1cm) rectangle (0.5cm,0.105cm); },
            legend to name=eff_role,
            xmin=1,xmax=11,
            ymin=0.001,ymax=1000,
            ytick = {0.001, 0.01, 0.1, 1, 10, 100, 1000},
            ymode = log,
            log origin=infty, 
            tick align=inside,
            ticklabel style={font=\Huge},
            every axis plot/.append style={line width = 2.5pt},
            every axis/.append style={line width = 2.5pt},
            ylabel={\textbf{\Huge Time (s)}}
            ]
\addplot[fill=c20] table[x=datasets,y=QTCS]{\queryt};
\addplot[fill=c15] table[x=datasets,y=QD-GNN]{\queryt};
\addplot[fill=c16] table[x=datasets,y=COCLEP]{\queryt};
\addplot[fill=c17] table[x=datasets,y=TransZero]{\queryt};
\addplot[fill=c18] table[x=datasets,y=CS-TGN]{\queryt};
\addplot[fill=c19] table[x=datasets,y=\model]{\queryt};

\legend{\small \texttt{QTCS}, \small \texttt{QD-GNN},\small \texttt{COCLEP}, \small \texttt{TransZero},\small \texttt{CS-TGN}, \small \texttt{\model}}
        \end{axis}
        \draw[red] (9.85,0.6) node[rotate=-90] {\bf \scriptsize OOM};
        \draw[red] (11.1,0.6) node[rotate=-90] {\bf \scriptsize OOM};
       \draw[red] (12.86,0.6) node[rotate=-90] {\bf \scriptsize OOM};
       \draw[red] (14.13,0.6) node[rotate=-90] {\bf \scriptsize OOM};
       
    \end{tikzpicture}
    \caption{Efficiency results of CS methods.}
    \label{fig:query_t}
\end{figure}
Figure \ref{fig:query_t} compares the running time of all methods in the search phase, with OOM denoting out-of-memory. Our model outperforms the baselines in terms of efficiency on most datasets, achieving over 100$\times$ speedup on the School dataset compared to \texttt{QTCS} due to our local search design. While \texttt{CS-TGN} is slightly more efficient than ours on the Brain and Patent datasets, it faces effectiveness issues and struggles to scale on large graphs.

$\bullet$ \textbf{Ablation Study.}
In this experiment, we evaluate the effect of three key modules by F1-score: Temporal Dynamics Modeling (TM), Node-Leiden Subgraph Alignment (NA), and Batch-Level Embedding Refinement (BR). We develop three variants by removing these modules from {\tt \model}, denoted as {\tt \model}-{\tt NA}, {\tt \model}-{\tt TM}, and {\tt \model}-{\tt BR-TM}. Table \ref{tab:ablation} shows the results: (1) The Temporal Dynamics Modeling module is crucial, as it captures temporal interaction information, which is vital for community search in temporal graphs.
(2) The Node-Leiden Subgraph Alignment module is also important for effectively grouping similar nodes into the same community.
\begin{table}[t]
    \caption{Ablation study.}
    \setlength{\abovecaptionskip}{-0.3cm}
    \setlength{\belowcaptionskip}{-0.3cm}
    \footnotesize
    \begin{tabular}{c|ccccc|c}
        \toprule 
    Models&School&Brain&Patent&arXivAI&arXivCS&+/-\\
        \midrule
        {\tt UTCS-BR-TM} &  0.9347 & 0.4497
&	0.2754& 0.1851
 & 0.2133&-13.66\%\\
        {\tt UTCS-TM} & 0.9289 & 0.4530&0.2812 & 0.1789 & 0.2203&-13.01\%\\
        {\tt UTCS-NA} & 0.9282 & 0.4892&0.2971 & 0.3266 & 0.2156&-1.93\%\\
        \midrule
    {\tt UTCS} & \textbf{0.9476} & \textbf{0.4894} & \textbf{0.3108} & \textbf{0.3285} & \textbf{0.2213} &-\\
        \bottomrule
    \end{tabular}
    \label{tab:ablation}
\end{table}

$\bullet$ \textbf{Case Study.}
\begin{figure}[t]
\vspace{-0.5cm}
    \centering

    \subfigure[COCLEP]{
        \centering
        \includegraphics[width=0.2\linewidth]{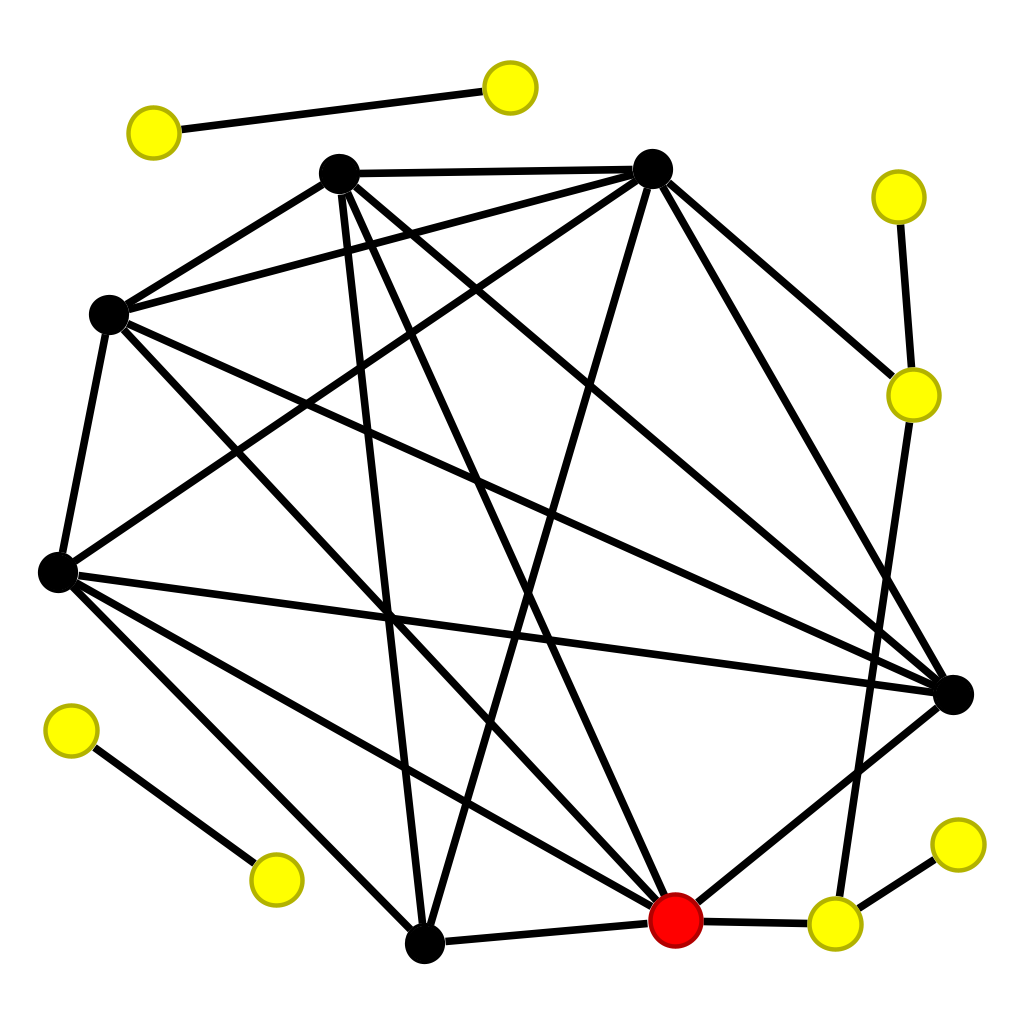}
        \label{fig:case_coclep}     
    }
    \subfigure[QTCS]{
        \centering
        \includegraphics[width=0.2\linewidth]{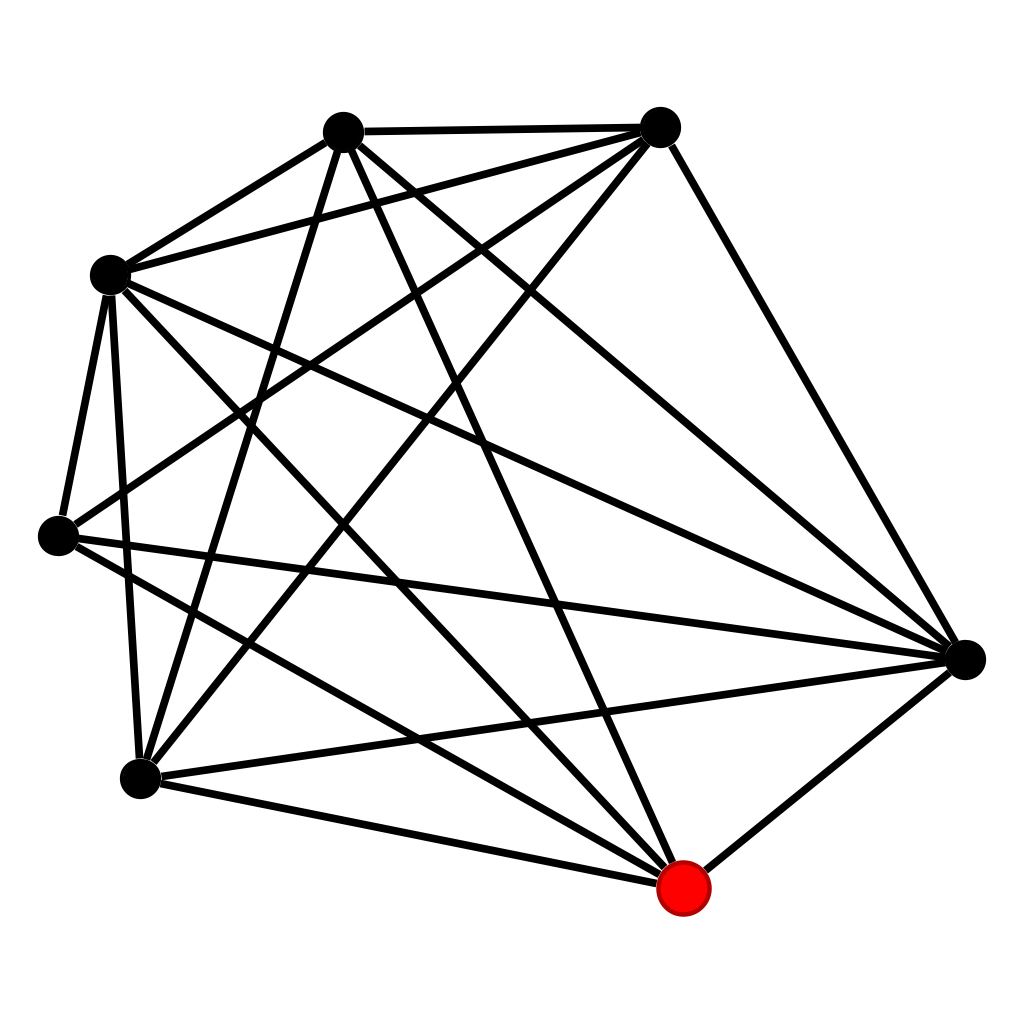}
        \label{fig:case_qtcs}     
    }
    \subfigure[\model]{
        \centering
        \includegraphics[width=0.2\linewidth]{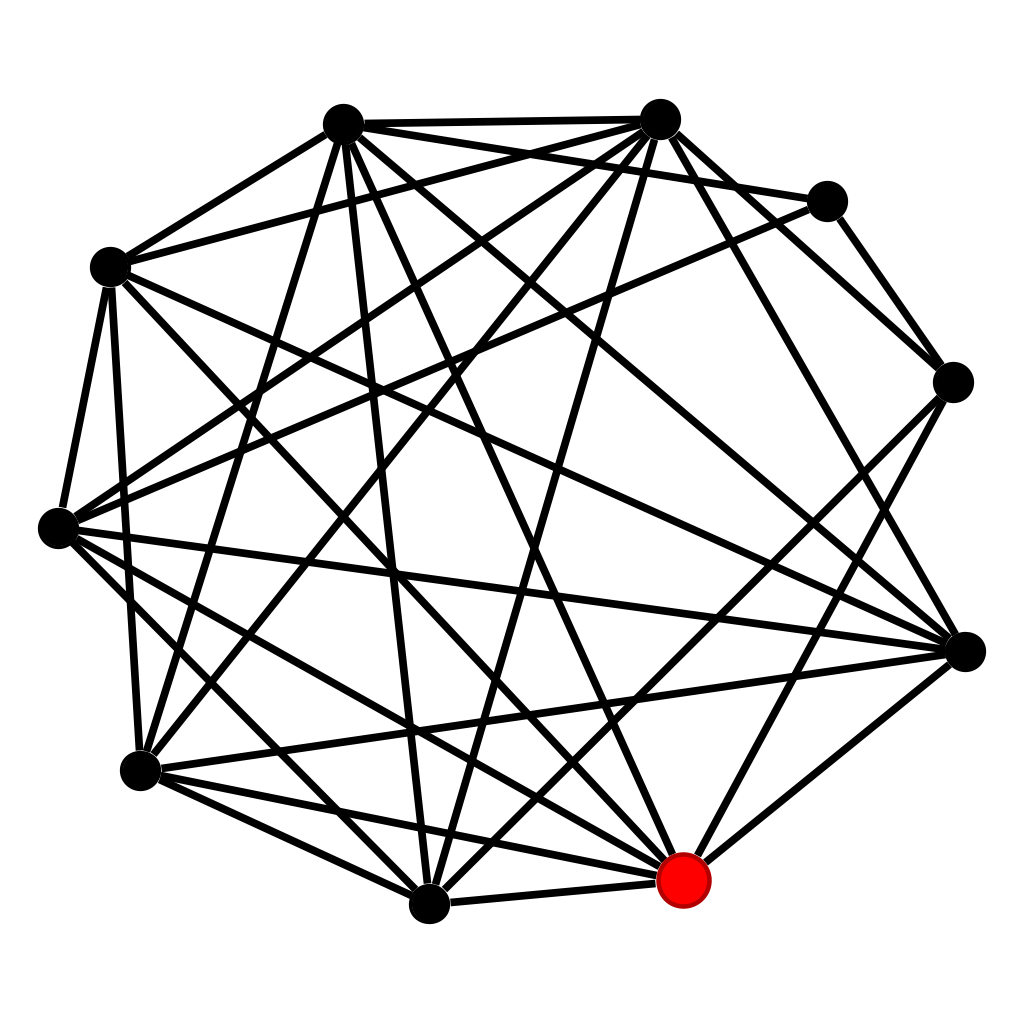}
        \label{fig:case_utcs}     
    }
    \subfigure[Ground-truth]{
        \includegraphics[width=0.2\linewidth]{ground_truth.png}
        \label{fig:case_gt}
    }	
    \caption{A case study on arXivAI dataset.}
    \label{fig:case study}
\end{figure}
We demonstrate the effectiveness of our method on the arXivAI dataset, where \texttt{QD-GNN} and \texttt{CS-TGN} encountered OOM issues. In Figure \ref{fig:case study}, the query node (red) is visualized alongside ground-truth community nodes (black) and non-community nodes (yellow). 
\texttt{COCLEP} misses several ground-truth nodes and includes many from other communities because it overlooks the temporal characteristics of the dataset. \texttt{QTCS} captures the temporal nature of the dataset but imposes strict constraints on temporal dimensions, missing nodes with weaker temporal associations. In contrast, our \texttt{UTCS} precisely identifies the ground-truth community.

\section{Conclusion and Future Work}
\label{sec:conclusion}
In this paper, we propose \model, the first unsupervised learning-based approach for the temporal CS problem. \model~consists of two phases. In the pre-training phase, temporal information is captured via the Hawkes process, while structural relationships are learned through the node-Leiden subgraph alignment and batch-level refinement modules. In the online phase, some subgraphs are extracted to narrow the search scope, and communities are predicted based on a new scoring mechanism. Experiments on five public datasets validate the method's effectiveness. 

Future work includes integrating traditional and learning-based methods to balance efficiency and accuracy, developing new metrics for unlabeled temporal graphs, and designing distributed algorithms to scale to massive temporal graphs.

\begin{acks}
This work was partially supported by NSFC under Grants 62302421 and 62302397, Basic and Applied Basic Research Fund in Guangdong Province under Grant 2023A1515011280, 2025A1515010439, Ant Group through CCF-Ant Research Fund, Shenzhen Research Institute of Big Data under Grant SIF20240004, the Guangdong Provincial Key Laboratory of Big Data Computing at The Chinese University of Hong Kong, Shenzhen, and the fund of the Laboratory for Advanced Computing and Intelligence Engineering (NO.2023-LYJJ-01-021).
\end{acks}

\bibliographystyle{ACM-Reference-Format}
\balance
\bibliography{ref}
\end{spacing}
\end{document}